\documentclass[twocolumn,preprintnumbers,amsmath,amssymb,showpacs,superscriptaddress]{revtex4}

\pdfoutput=1 

\usepackage{graphicx}
\usepackage{bm}
\usepackage{dcolumn}
\usepackage{amssymb}
\usepackage{amsmath}
\usepackage{amsfonts}
\usepackage{color}
\usepackage{epsfig}
\usepackage{subfigure}
\newcommand{\ket}[1]{|#1\rangle}

\begin{document}
\title{Self-consistent $\textbf{k}\cdot \textbf{p}$ calculations for gated thin layers of 3D Topological Insulators 
}
\author{Yuval Baum} 
\affiliation{Department of Condensed Matter Physics, Weizmann Institute of Science, Rehovot 76100, Israel}
\author{Jan B{\"o}ttcher}
\affiliation{Institut f{\"u}r Theoretische Physik und Astrophysik (TP4), Universit{\"a}t W{\"u}rzburg}
\author{Christoph Br{\"u}ne}
\affiliation{Physikalisches Institut (EP3), Universit{\"a}t W{\"u}rzburg, Am Hubland, 97074 W{\"u}rzburg, Germany}
\author{Cornelius Thienel}
\affiliation{Physikalisches Institut (EP3), Universit{\"a}t W{\"u}rzburg, Am Hubland, 97074 W{\"u}rzburg, Germany}
\author{Laurens W.~Molenkamp}
\affiliation{Physikalisches Institut (EP3), Universit{\"a}t W{\"u}rzburg, Am Hubland, 97074 W{\"u}rzburg, Germany}
\author{Ady Stern} 
\affiliation{Department of Condensed Matter Physics, Weizmann Institute of Science, Rehovot 76100, Israel}
\author{Ewelina M.~Hankiewicz}
\affiliation{Institut f{\"u}r Theoretische Physik und Astrophysik (TP4), Universit{\"a}t W{\"u}rzburg}

\begin{abstract}
Topological protected surface states are one of the hallmarks of three-dimensional topological insulators.
In this work we theoretically analyze the gate-voltage-effects on a quasi-3D layer of HgTe. We find that while the gapless surface states dominate the transport, as an external gate voltage is applied, the existence of bulk charge carriers is likely to occur. We also find that due to screening effects, physical properties that arise from the bottom surface are gate-voltage independent. Finally, we point out the experimental signatures that characterize these effects.
\end{abstract}
\pacs{03.65.Vf, 72.20.-i, 72.25.-b, 72.20.My, 73.40.-c}
\maketitle

\section{Introduction}
Since their prediction in 2005, topological insulators have become one of the main themes in condensed matter physics \cite{mele,FKM,moore,bhz,hassan,koning}. Protected gapless surface states are among the manifestations of these materials. These surface states have drawn a great deal of interest, both as a platform for the investigation of fundamental physics and due to their applications potential.
Their equilibrium properties have been extensively studied in recent years using ARPES and STM techniques \cite{Chen,Hsieh,Hsieh2,Hanaguri}, mostly in narrow gap materials such as Bi$_2$Te$_3$ and Bi$_2$Se$_3$. However, the strong intrinsic doping in these materials leads to a bulk dominated conductance, which makes the study of their transport properties difficult.
Following theoretical predications \cite{fu,Dai,saad}, it has been proven in recent experiments \cite{brune} that quasi-3D layers of HgTe (between $40$ to $200$ $nm$ thick) which are grown epitaxially on a CdTe substrate become three-dimensional topological insulators. These topological insulators may host high mobility surface states that dominate the transport whenever the chemical potential is tuned to the bulk gap. In thin samples (less than $40\,nm$), the two surface states hybridize and the surfaces will become gapped. 

The topological surface states in HgTe arise as a consequence of an inverted band structure in the bulk. Undoped HgTe is a semi-metal, hence, the role of the CdTe substrate is crucial in order to get a bulk insulating HgTe. Due to a $0.3\%$ lattice mismatch between the materials, the CdTe applies strain on the HgTe layer. It turns out that this strain opens a small gap \cite{fu,Roman} between the light and the heavy holes bands. This gap hosts the metallic topological surface states.
When the thickness of the sample exceeds $200\,nm$, a relaxation of the elastic energy will lead to dislocations and the HgTe will restore its semi-metallic nature.
  
Motivated by experiments \cite{brune}, in this work, we analyzed gate-voltage effects on a quasi-3D layer of HgTe within the framework of the $\textbf{k}\cdot\textbf{p}$ theory and the Hartree approximation \cite{Novik}. Additionally, we will try to relate our theoretical findings to the experimental data.

The samples in the relevant experiments \cite{brune}, contain a quasi-3D HgTe layer which is grown on top of a CdTe substrate. Some of the samples may also have a CdTe (CdHgTe) cap layer on top of the HgTe layer. Then, the sample is structured into a Hall bar geometry. The entire sample is covered by a multilayer insulator, and a top gate electrode is deposited on top of it. 
Our model describes the band structure, near the $\Gamma$ point, of these HgTe samples. In particular, we are interested in the interplay between the internal electrostatic effects and the external gate-voltage effects. 
   
\section{The band structure model}
HgTe has a zinc-blende structure. Its electronic properties are determined by the band structure near the Fermi surface at the $\Gamma$ point.
Our starting point is the usual six-band (we neglected the spin-orbit split off $\Gamma_7$ band) basis set $\Big(\ket{\Gamma_6,1/2},\,\ket{\Gamma_6,-1/2}\Big)$, which form the conduction band in normal materials, and $\Big(\ket{\Gamma_8,3/2},\,\ket{\Gamma_8,1/2},\,\ket{\Gamma_8,-1/2},\,\ket{\Gamma_8,-3/2}\Big)$ which form the holes bands in normal materials. 
In HgTe there is a strong mixing between electronic states in the bottom of the conduction band and electronic states at the top of the valence band.
This physics near the $\Gamma$ point is captured by the six-band Kane Hamiltonian, $H_0$, \cite{kane} in the axial approximation (see appendix \ref{ap1}).
  
Next, in order to open an insulating gap at the Fermi energy we considered a uniaxial strain along the (001) direction. The effect of strain can be taken into consideration by introducing an additional strain induced Hamiltonian, $H_s$, according to the Bir and Pikus formalism \cite{bir} (see appendix \ref{ap1}).  
Finally, in order to describe a system with a finite size in the $z$-direction and in order to consider electrostatic effects, we first perform a lattice regularization (which is valid in the vicinity of the $\Gamma$ point): $k_i\to \frac{1}{a}\sin{(ak_i)}$ and $k_i^2\to \frac{2}{a^2}[1-\cos{(ak_i)}]$, where $a$ is the lattice constant.
Next, by performing an inverse Fourier transform in the $z$-direction, we get an effective one dimensional tight-binding model in the $z$-direction:
\begin{align}\label{TB}
H(\textbf{k}_{\parallel})=\sum\limits_{n=1}^{N_z}&C_{i,n}^{\dagger}(\textbf{k}_{\parallel})\hat{A}_{ij}(\textbf{k}_{\parallel})C_{j,n}(\textbf{k}_{\parallel})+\\ \nonumber
&\Big[C_{i,n}^{\dagger}(\textbf{k}_{\parallel})\hat{T}_{ij}(\textbf{k}_{\parallel})C_{j,n+1}(\textbf{k}_{\parallel})+h.c.\Big]
\end{align}
where $C_{i,n}^{\dagger}(\textbf{k}_{\parallel})$ is the creation operator of an $i$'s band electron in layer $n$ and with in-plane momentum $\textbf{k}_{\parallel}$. $N_z$ is the number of layers in the $z$-direction. The coefficients $\hat{A}_{\textbf{k}_{\parallel}},\,\hat{T}_{\textbf{k}_{\parallel}}$ are $6\times6$ matrices that can be extracted from the six-band Kane Hamiltonian. Their exact form appears in appendix \ref{ap2}. Eq.~(\ref{TB}) can be solved numerically for each value of $\textbf{k}_{\parallel}$ to produce a set of $6N_z$ eigenvalues, $\mathcal{E}_{n,\textbf{k}_{\parallel}}$, and $6N_z$ eigenstates $\psi_n(\textbf{k}_{\parallel},z)$.
The charge density is now defined as:
\begin{equation}\label{charge}
\rho(z)=-e\sum\limits_{n=1}^{6N_z}\int{\frac{d^2k}{(2\pi)^2}|\psi_n(\textbf{k}_{\parallel},z)|^2f_{FD}(\mathcal{E}_{n,\textbf{k}_{\parallel}}-\mu)}-\rho_0
\end{equation}   
where $\mu$ is the chemical potential, $\rho_0$ is a neutralizing background and $f_{FD}$ is the Fermi-Dirac distribution. The internal potential due to this electronic density is derived via the Poisson equation: 
\begin{equation}\label{Poisson}
\partial_{zz}\phi_H(z)=-\frac{\rho(z)}{\epsilon_0\epsilon_r}
\end{equation} 
where $\phi_H(z)$ is the Hartree-potential, and $\epsilon_r\approx21$ is the relative dielectric constant of HgTe.

\begin{figure}[h]
\centering
 {
   \includegraphics[scale =0.45] {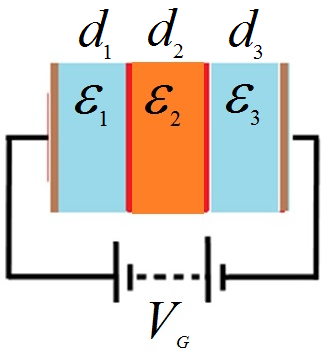} 
 }
\caption{Capacitor model: charge will accumulate in the interfaces between the different dielectric materials.}
\label{Fig_cap}
\end{figure}

\begin{figure}[h]
\centering
 {
   \includegraphics[scale =0.38] {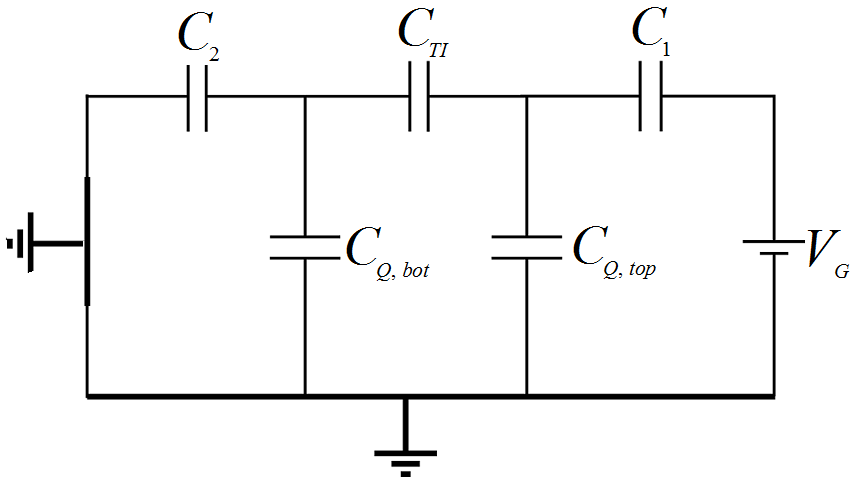} 
 }
\caption{Capacitor model for a grounded sample. Here $C_1$ represents the geometrical capacitance of the layers between the gate and the sample, $C_2$ represents the geometrical capacitance of the layers between the sample and the substrate, $C_{TI}$ represents the geometrical capacitance of the topological insulator layer and $C_{Q, top},\,C_{Q, bot}$ represent the quantum capacitance of the top and the bottom surface states.}
\label{cap_net}
\end{figure}

Now, we would like to consider the external gate voltage. We assume that the components of the sample behave as a series of capacitors. 
As demonstrated in Fig.~(\ref{Fig_cap}), although the components of the sample are connected (and not separated by metallic plates like in a standard series of capacitors), charge will accumulate in the interfaces between the different dielectric materials, and inside each component an electric field will form. This field is proportional to the capacitance of each component. 
Hence, we first assume that the voltage-drop on each component is proportional to its geometrical capacitance. Using this, we can estimate $\tilde{V}_G$, the voltage-drop on the HgTe layer only due to the external gate-voltage $V_G$.
It should be pointed out that in \cite{brune} a different experimental configuration is used, where the topological insulator layer (HgTe) is grounded. In this case, the components of the sample do not behave as capacitors in series. By considering the quantum capacitance of the top and the bottom surface states, the electrostatics of the sample, without the Hartree, is captured by the effective capacitor network in Fig.~(\ref{cap_net}) in contrast to Fig.~(\ref{Fig_cap}).
Now, the portion of the gate voltage that drops on the HgTe layer, $\tilde{V}_G$, may be different. In general, for a given experimental configuration, the value of $\tilde{V}_G$ should be extracted from the equivalent circuit of the experimental setup. In this paper, we will not limit ourselves to a specific experimental setup. All the results in the paper are general, up to the fact that for a given gate voltage, the actual value of $\tilde{V}_G$ will depend on the specific details of the experimental setup.
In realistic samples, the voltage drop across the HgTe layer will be several percent of the total gate voltage
We also assumed that the gate-voltage drops linearly on each component of the sample. Therefore, the gate potential within the HgTe layer can be approximated by : $\phi_G(z)=\frac{\tilde{V}_G}{L_z}z$, where $L_z=aN_z$ is the length of the sample in the $z$-direction. Now, the energy due to the two potentials is given by $V(z)=-e[\phi_G(z)+\phi_H(z)]$. Treating this energy as an on-site energy in the tight-binding model, Eq.~(\ref{TB}) is modified:
\begin{align}\label{TB2}
&H(\textbf{k}_{\parallel},V_G)=\\ \nonumber
&\sum\limits_{n=1}^{N_z}C_n^{\dagger}\Big(\hat{A}_{\textbf{k}_{\parallel}}+V(z_n)\hat{I}\Big)C_n+\Big[C_n^{\dagger}\hat{T}_{\textbf{k}_{\parallel}}C_{n+1}+h.c.\Big]
\end{align}
where $\hat{I}$ is a $6\times6$ unit matrix. Eq.~(\ref{TB2},\ref{charge},\ref{Poisson}) are coupled and must be solved in a self-consistent manner. In order to do so, the value of the chemical potential, $\mu$, must be determined. We assume a thermodynamical equilibrium, hence, the two surfaces equilibrate when the gate voltage is varied and the chemical potential is a constant throughout the sample. We determine the value of $\mu$ in such a way that any change in the gate voltage $\Delta V_G$ will produce a change in the system charge according to $\Delta Q=C\Delta V_G$, where $C$ is the geometrical capacitance of the sample.
In general, the capacitance of the sample depends on the experimental setup, as is evident from Figs.~(\ref{Fig_cap},\ref{cap_net}). It is a combination of both the geometrical capacitance and the quantum capacitance of the two dimensional surface states.
Here, we neglect the effect of quantum capacitance on the sample capacitance. This assumption is justified as long as the parameter $(2\alpha)^{-1}(v_F/c)(\epsilon_{ox}/k_\texttt{F} d_{ox})$ is small. Where $\epsilon_{ox}$ and $d_{ox}$ are the relative permittivity and the thickness of the oxide layer, $c$ is the speed of light, $\alpha$ is the fine structure constant and $v_F$ is the Fermi-velocity of the surface spectrum. This condition is met as long as the chemical potential is not at the vicinity of the Dirac-point. See appendix ~\ref{qCap} for the detailed calculation. Note that as long as this condition is met, the total charge on the sample will not depend on the specifics of the experimental setup.
In general, the self consistent calculation should incorporate both the band structure calculation and the electrostatics arising from the appropriate equivalent circuit. However, as we show in appendix~\ref{qCap}, in these types of samples, the quantum capacitance of the surface states is much larger than the typical geometrical capacitance of the sample. Hence, using a specific scheme with a specific equivalent circuit will not change the qualitative results. The total charge will not change, and the only difference will be a rescaling of voltage that drops on the topological insulator layer. To be specific, throughout this paper we will use the scheme of capacitors in series. 
Using that scheme, a self-consistent solution to Eq.~(\ref{TB2},\ref{charge},\ref{Poisson}) can be obtained numerically in order to find the band-structure of HgTe in the presence of a gate voltage.
\section{Band Model Results}
\begin{figure}[b]
\centering
\subfigure{
   \includegraphics[scale =0.45] {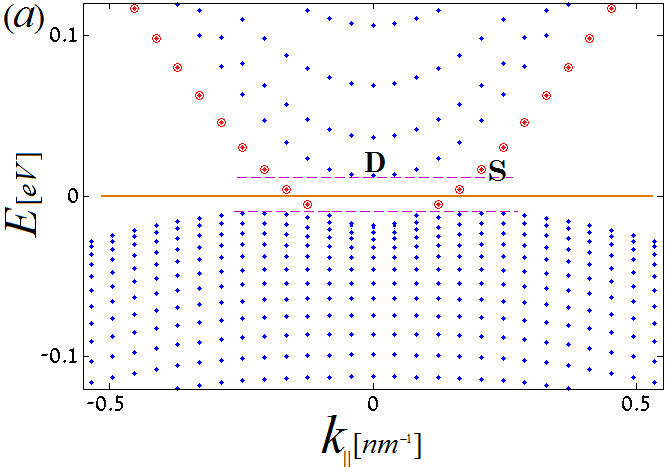}
   \label{Fig1a}
 }
\subfigure{
   \includegraphics[scale =0.45] {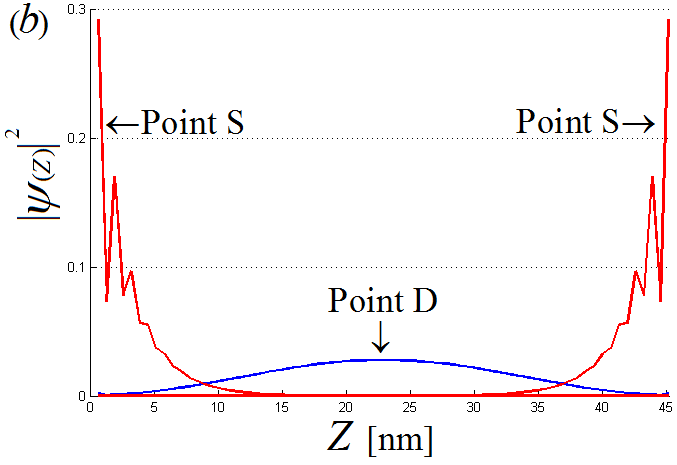}
   \label{Fig1b}
 }
\caption{(a) The band structure near the $\Gamma$ point for $V_G=V_0$. All the points are doubly degenerate. The orange line is the chemical potential, and the purple lines denotes the bulk gap. The red points belong to the surface states. (b) The absolute square of the wave functions in the selected points $S$ and $D$. Clearly, the two states at point $S$ are the top and bottom surface states.}
\end{figure}

We will present both the band structure near the $\Gamma$ point and selected wave functions.
All the following results are calculated for $70$ layers in the $z$-direction ($N_z=70$). This corresponds to a sample thickness of $\sim45\,nm$. We verified that the results
do not depend qualitatively on the number of grid points in this direction. The following results will not change qualitatively as long as the sample thickness is less than $120\,nm$.   
Points in the band structure which belong to the surface (their wave functions decay exponentially to the bulk) will appear in red, while bulk points will appear in blue. The orange line denotes the chemical potential.
First we must choose a reference point. There is a gate voltage at which the two surface states (top and bottom) are degenerate. We will refer to this point as the symmetric point. We denote its corresponding gate-voltage by $V_0$.
We also know that at the symmetric point the chemical potential should lie in the bulk gap. The exact location of the chemical potential could be determined by measuring the exact electron density in the symmetric point, which experimentally can be extracted from Hall measurements. To a good approximation, we will fix the chemical potential in the symmetric point to the middle of the bulk gap.
The band structure near the $\Gamma$ point and the wave functions in selected points appear in fig.~(\ref{Fig1a},\ref{Fig1b}).
Clearly, the two surface states are degenerate and well localized near the surfaces for energies in and above the bulk gap. For energies near the bottom of the conduction band, the surface sub-bands are separated (in momentum) from the bulk sub-bands. This reduces the hybridization between the surface and the bulk states, and hence, the surface sub-bands remain exponentially localized near the surfaces.
For energies below the bulk gap (the valence band) the surface sub-band are not separated (in momentum) from the bulk sub-bands, and the surface sub-bands significantly hybridize with the bulk states.

Another prominent feature in the band structure is the relatively large level spacing of the conduction sub-bands, which is a result of the small effective mass of the conduction electrons ($m_e^*\approx 0.03m_0$). Clearly, the wave functions of these sub-bands (point $D$) are "bulk" wave functions and their center of charge is located in the middle of the bulk. 

\begin{figure}[b]
\centering
\subfigure{
   \includegraphics[scale =0.44] {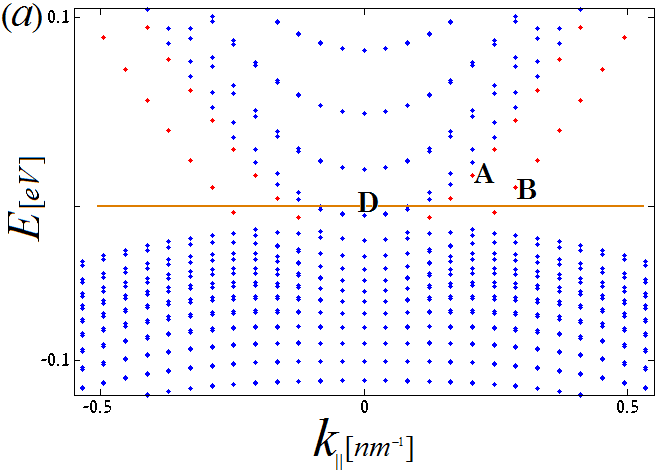} 
   \label{Fig2a}
 }
\subfigure{
   \includegraphics[scale =0.44] {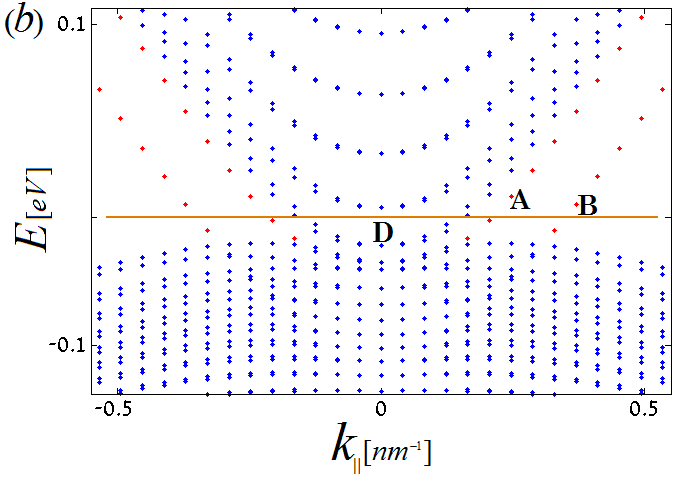}
   \label{Fig2b}
 }
 \subfigure{
   \includegraphics[scale =0.44] {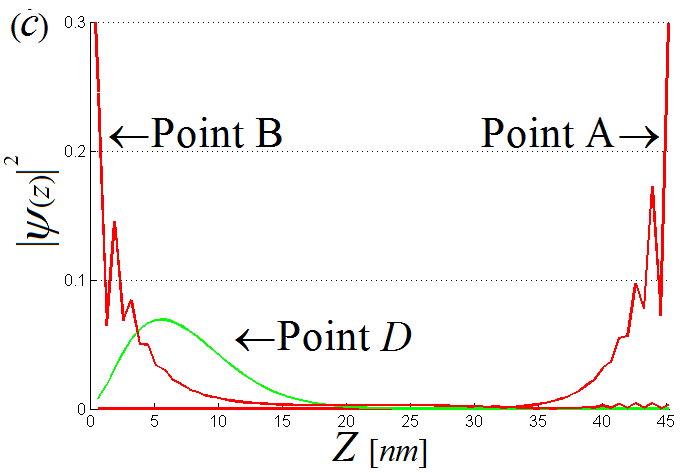}
   \label{Fig2c}
 }
\caption{The band structure near the $\Gamma$ point for $V_G=V_1\,(V_0<V_1)$ (a) and $V_G=V_2\,(V_1<V_2)$ (b). The orange line is the chemical potential, and the red points belong to the surface states. (c) The absolute square of the wave functions in the selected points $A$, $B$ and $D$ in the first band structure. The center of charge of bulk states moves towards the top surface.}
\end{figure}

Now we will examine the effect of positive gate voltages.
The band structure near the $\Gamma$ point for two different positive gate voltages ($V_0<V_1<V_2$) appears in Fig.~(\ref{Fig2a},\ref{Fig2b}). Also here, the orange line denotes the chemical potential, and the red points belong to the surface states.

\begin{figure}[h]
\centering
 {
   \includegraphics[scale =0.42] {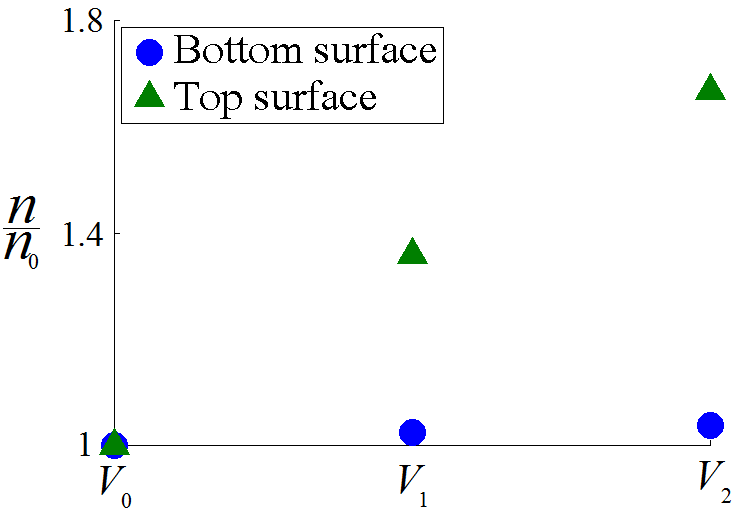} 
 }
\caption{Normalized densities of the top and bottom surface states for gate voltages $V_0$, $V_1$ and $V_2$. Clearly, while the density of the top surface increases, the density of the bottom surface remains almost unchanged.}
\label{den}
\end{figure}

Clearly, the surface sub-bands are not degenerate in energy anymore. By comparing the relative shift (compare to the symmetric point) of the surface sub-bands with respect to the chemical potential, we conclude that while the density of the top surface increases, the density of the bottom surface remains almost unchanged. The densities of the top and bottom surface states as extracted from the band structures in Fig.~(\ref{Fig2a},\ref{Fig2b}) appear in Fig.~(\ref{den}). It is also clear that the bottom of the first conduction sub-bands becomes populated, and the center of charge of these states (point $D$) moves towards the top surface.

After examining a large variety of positive gate voltages, we can identify three prominent features:
\begin{enumerate}  
\item For a large range of gate voltages ($V_0<V_G$), the chemical potential crosses two (well defined) surface sub-bands and two (single mode) conduction sub-bands. All these bands are well separated from the other bulk sub-bands.
\item As the gate voltage varies (becomes more positive) the top surface sub-band becomes more and more populated, while the occupation of the bottom surface sub-band is almost unchanged. This scenario is different from electron-compressibility measurements of two-dimensional systems, such as \cite{QC}, where due to the lack of bulk carriers, the bottom 2DEG is also affected.
\item As the gate voltage varies, the first conduction sub-band becomes more and more populated, and its center of charge moves towards the top surface.
\end{enumerate} 
The last two features are a result of a screening mechanism. Both the top-surface and the bulk screen the charge on the gate electrode. Hence we expect that physical quantities that arise from the bottom surface state will be gate independent. Moreover, we expect that for a large range of gate voltages and at low temperatures the transport in the system will be described by three decoupled quasi-$2D$ charge carriers. The exact range of gate voltages strongly depends on the specifics of the sample. From the results of our theoretical description, we expect to observe this phenomenon up to approximately five volts above the symmetric-point voltage.

Now we turn to negative gate voltages. For large negative gate voltages the chemical potential enters the valence band, where the surface states are completely hybridized. However, for moderate negative gate voltages the situation is similar to the positive gate-voltage case. 
The band structure near the $\Gamma$ point for a moderate negative gate voltage ($V<V_0$) appears in Fig.~(Fig6). Also here, the orange line denotes the chemical potential, the red points belong to the surface states and the green points belong to the first sub-band of the valence band. The wave functions of the selected points are similar to Fig.~(\ref{Fig2c}). 
\begin{figure}[h]
\centering
 {
   \includegraphics[scale =0.45] {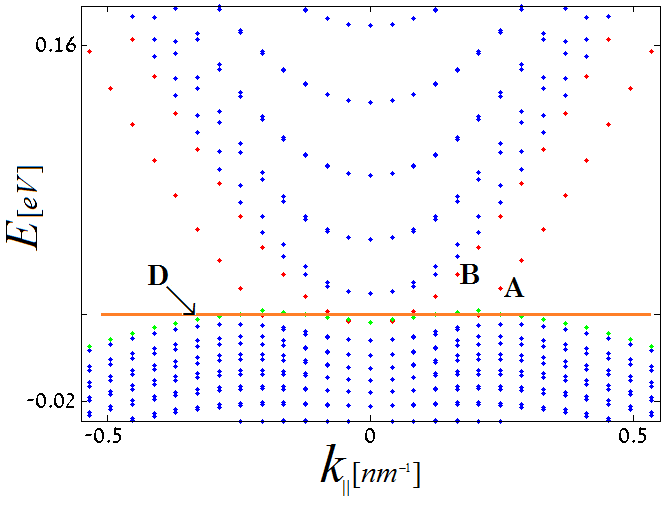} 
 }
\caption{The band structure near the $\Gamma$ point for $V_0>V_G$. The orange line is the chemical potential, and the red points belong to the surface states and the green points belong to the first sub-band of the valence band. The absolute square of the wave functions in the selected points $A$, $B$ and $D$ in the band structure are similar to wave functions in fig.~\ref{Fig2c}}.
\label{Fig6}
\end{figure} 

Also in this case the chemical potential crosses three sub-bands, two surface states and the first sub-band of the valence band. As in the positive gate voltage case, the density of the bottom surface is almost unchanged. However, the density of the top surface reduces and some bulk hole-states become populated. Also here the center of charge of these hole-states (point $D$) moves towards the top surface. Hence, a similar screening mechanism occurs also for negative voltages and as long as the chemical potential is at the top of the valence band we expect to observe a similar behavior to positive voltage case. Nevertheless, we expect to observe this behavior only for a narrow range of gate voltages, since as the gate voltage increases, the chemical potential will decrease to an energy range where the surface states are not well defined and many bulk sub-bands are populated.
Therefore we expect to observe an asymmetry with respect to the gate voltage sign.

\section{Experimental Signatures}
It is instructive to examine how transport in a quasi-3D slab of strain HgTe depends on external probes such as gate voltage and magnetic field. 
The most prominent feature of the results above is the fact that for a large range of gate voltages, the chemical potential crosses three distinct sub-bands. Therefore we expect that for a large range of gate voltages and at low temperatures the transport in the system will be described by three decoupled quasi-$2D$ charge carriers.
In particular, in the presence of a magnetic field the resistivity tensor of a charge carrier with charge $q_i$, density $n_i$ and mobility $\mu_i$ is given by:
\begin{equation}\nonumber
\hat{\rho}_i=\begin{pmatrix} 1/\sigma_i & R_iB \\
-R_iB & 1/\sigma_i \\
\end{pmatrix}
\end{equation}
where $\sigma_i=|q_i|\mu_i n_i$ and $R_i=(q_i n_i)^{-1}$.

For multiple charge carriers the total resistivity tensor of the system is given by:
\begin{equation}\label{LL}
\hat{\rho}=\Big(\sum\limits_{i}(\hat{\rho}_i)^{-1}\Big)^{-1}
\end{equation}
In the presence of a low magnetic field, each $\sigma_i$ may experience Shubnikov-de Haas (SdH) oscillations \cite{SdH} of the form:
\begin{equation}\label{LL}
\sigma_i(B)=\frac{\sigma_i(B=0)}{1+\phi_i^2}\,\Big[1-\frac{2\phi_{c,i}^2}{1+\phi_{c,i}^2}\,\frac{x_i}{\sinh{(x_i)}}\,e^{-\pi/\phi_c}\,\cos{\Big(\frac{F_i}{B}\Big)}\Big]
\end{equation}
where $F_i=2\pi\Phi_0n_i$, $x_i=\frac{2\pi^2k_BT}{\hbar\omega_{c,i}}$, $\phi_i=\mu_i\,B$, $\phi_{c,i}=\mu_{c,i}\,B$, $\omega_{c,i}$ is the cyclotron frequency and $\Phi_0$ is a flux quantum. The quantity $\mu_{c,i}$ is related to $\mu_{i}$ by $\frac{\mu_{c,i}}{\mu_{i}}=\frac{\tau_c}{\tau}$, where $\tau_c$ is the cyclic relaxation time corresponding to the dephasing of the Landau state, which may be quite different from the transport relaxation time $\tau$.

For the general case where the density and mobility of each charge carrier are different, both the diagonal and the off-diagonal parts of the resistivity tensor will contain oscillatory terms and their magnitude will depend on the magnitude of the magnetic field. Moreover, since the frequency of these oscillations depend only on the density of the charge carriers, we expect that the frequency of the oscillatory components that arise from the bottom surface will be gate voltage independent.
As the last experiments show \cite{brune}, the topological surface states in these HgTe samples may have high mobility at low temperatures, hence we expect low-field magneto-transport measurements to show the following signatures:
\begin{enumerate}  
\item Strong magneto-resistance, $R_{xx}$ will increase as a function of magnetic field.
\item Oscillation as a function of $1/B$, both in $R_{xx}$ and in $R_{xy}$.
\item Existence of a gate voltage independent component. In particular, if the bottom surface has the highest mobility, we expect that for low fields, the frequency of the oscillations to be nearly gate independent.
\end{enumerate} 
It should be pointed out that the analysis above is valid only in the low-field regime. In high-field, the two Dirac surface states form Landau-levels. Hence, our band model is not adequate in this regime. Indeed, recent experiments show \cite{brune} a high-field behavior that can not be explained by the above model.
\section{Summary}
To summarize, we considered here the effect of an external gate voltage on the electronic and on the transport properties of a quasi-3D layer of strained HgTe. 
We assumed that these properties arise solely from the band structure near the Fermi surface at the $\Gamma$ point. We analyzed the band structure within the framework of the $\textbf{k}\cdot\textbf{p}$ theory, and we included electrostatic effect within the framework of the the Hartree approximation.

We found in our model that for a large range of gate voltages, the chemical potential crosses three decoupled states, two well defined surface states and a single mode bulk state.
We also found that as the gate voltage varies, both the top-surface state and the bulk states screen the charge on the gate electrode, and therefore the bottom surface and in particular its density become gate-voltage independent.
Finally, we conclude that for a large range of gate voltages and at low temperatures the transport in these systems may be described by three decoupled quasi-$2D$ charge carriers, and we pointed out the main experimental signatures of such a scenario.

\begin{acknowledgements}
The authors thanks the European Research Council under the 
European Community's Seventh Framework Program, Microsoft Station Q, the US-Israel BSF and the Minerva Foundation for financial support. E.~M.~H. and J.~B. were supported by the German Research
Foundation DFG grant HA 5893/4-1 within SPP 1666.
L.~W.~M.~ acknowledges support by the German Research Foundation (DFG-JST joint research program "Topological
Electronics"), and EU ERC-AG program (project 3-TOP).
\end{acknowledgements}

\appendix

\section{$\textbf{k}\cdot \textbf{p}$ and strain Hamiltonians}\label{ap1}
Before including electrostatic effects, the approximated Hamiltonian (near the $\Gamma$ point) of strained HgTe is given by: $H=H_0+H_{\mbox{st}}$, where $H_0$ is the $\textbf{k}\cdot \textbf{p}$ Hamiltonian and $H_{\mbox{st}}$ is the strain Hamiltonian.
The $\textbf{k}\cdot \textbf{p}$ Hamiltonian is given by the six-band Kane Hamiltonian \cite{kane}.
\begin{widetext}
In the axial approximation, the form of $H_0$ in the basis $\Big(\ket{\Gamma_6,1/2},\,\ket{\Gamma_6,-1/2},\,\ket{\Gamma_8,3/2},\,\ket{\Gamma_8,1/2},\,\ket{\Gamma_8,-1/2},\,\ket{\Gamma_8,-3/2}\Big)^{\dagger}$ is:
\begin{equation}\label{kp}
H_0=\begin{pmatrix} 
T& 0 & -\frac{1}{\sqrt{2}}Pk_+ & \sqrt{\frac{2}{3}}Pk_z & \frac{1}{\sqrt{6}}Pk_- &0 \\
0& T & 0 &-\frac{1}{\sqrt{6}}Pk_+ &\sqrt{\frac{2}{3}}Pk_z&\frac{1}{\sqrt{2}}Pk_-\\
-\frac{1}{\sqrt{2}}Pk_- & 0 & U+V & 2\sqrt{3}\gamma Bk_{-}k_{z}&\sqrt{3}\gamma Bk_{-}^2&0\\
\sqrt{\frac{2}{3}}Pk_z &-\frac{1}{\sqrt{6}}Pk_- &2\sqrt{3}\gamma Bk_{+}k_{z}& U-V &0&\sqrt{3}\gamma Bk_{-}^2\\
\frac{1}{\sqrt{6}}Pk_+ & \sqrt{\frac{2}{3}}Pk_z& \sqrt{3}\gamma Bk_{+}^2 & 0 &U-V& -2\sqrt{3}\gamma Bk_{-}k_{z} \\
0& \frac{1}{\sqrt{2}}Pk_+ & 0 &\sqrt{3}\gamma Bk_{+}^2& -2\sqrt{3}\gamma Bk_{+}k_{z} &U+V\\\end{pmatrix}
\end{equation}
where $k_{\pm}=k_x\pm ik_y$, $B=\hbar^2/(2m_0)$, $P=\sqrt{BE_P}$, 
$T=E_g+B(k_x^2+k_y^2+k_z^2)$, $U=-B\hat{\gamma}(k_x^2+k_y^2+k_z^2)$ and $V=-B\gamma(k_x^2+k_y^2-2k_z^2)$.
Where for HgTe: $\gamma=0.9$, $\hat{\gamma}=4.1$, $E_g=-0.3eV$ and $E_P=18.8eV$ \cite{Novik}. 
\end{widetext}
The subspace of $\Big(\ket{\Gamma_8,3/2},\,\ket{\Gamma_8,-3/2}\Big)^{\dagger}$ gives the heavy holes (HH) sub-bands that belong to the valence band.
The subspace of $\Big(\ket{\Gamma_6,1/2},\,\ket{\Gamma_6,-1/2},\,\ket{\Gamma_8,1/2},\,\ket{\Gamma_8,-1/2}\Big)^{\dagger}$ gives the electrons sub-bands that belong to the valence band and the light holes (LH) sub-bands that belong to the conduction band.

The strain Hamiltonian, $H_{st}$, is obtained by the Bir-Pikus formalism \cite{bir}. The origin of the strain in the HgTe is the $0.3\%$ lattice mismatch between the HgTe layer and the CdTe substrate. The Bir-Pikus formalism assumes that the components of the strain tensor $\epsilon_{ij}$ transform under rotations similar to the tensor $k_ik_j$. Eq.~(\ref{kp}) is a general form (up to second order in $\textbf{k}$) that respect the lattice symmetry. Therefore, a Hamiltonian that respects the zinc-blende lattice symmetry and is proportional to the component of the strain tensor is derived from Eq.~(\ref{kp}) by the substitution $k_ik_j\to \epsilon_{ij}$ followed by the replacement of the band structure parameters by the deformation potentials as follows \cite{Caro}:
\begin{align}
\frac{\hbar^2}{2m_0}\to C \\ \nonumber
\frac{\hbar^2\hat{\gamma}}{2m_0}\to -a\\ \nonumber
\frac{\hbar^2\gamma}{m_0}\to -b \\ \nonumber
\end{align}
where $C$ and $a$ are the hydrostatic deformation potentials and $b$ is the the uniaxial deformation potential.
For a uniaxial strain along the (001) direction, $\epsilon_{ij}$ is diagonal with $\epsilon_{xx}=\epsilon_{yy}=-\frac{1}{2x}\epsilon_{zz}\equiv\epsilon$, where $\epsilon=0.003$ is given by the lattice mismatch, and $x=\frac{C_{12}}{C_{11}}$, where $C_{12}$ and $C_{11}$ are elastic stiffness constants.
\begin{widetext}
Hence, the strain Hamiltonian, $H_{st}$, that is obtained by the Bir-Pikus formalism:
\begin{equation}\label{Hs}
H_{st}=\begin{pmatrix} 
T_{\epsilon}& 0 & 0 & 0 &0 &0 \\
0& T_{\epsilon} & 0 & 0 &0 &0\\
0 &0 & U_{\epsilon}+V_{\epsilon} &0&0&0\\
0 &0 &0& U_{\epsilon}-V_{\epsilon} &0&0\\
0 & 0& 0 & 0 &U_{\epsilon}-V_{\epsilon}& 0 \\
0& 0 & 0 &0&0 &U_{\epsilon}+V_{\epsilon}\\\end{pmatrix}
\end{equation}
where for HgTe: 
$T_{\epsilon}=0.4\epsilon(1-x)\,[eV]$, $U_{\epsilon}=-6.8\epsilon(1-x)\,[eV]$ and $V_{\epsilon}=-1.5\epsilon(1+2x)\,[eV]$, with $\epsilon=0.003$ and $x=0.68$ \cite{Adachi}.
\end{widetext}

\section{Tight binding Hamiltonian}\label{ap2}
In general, the translational symmetric Hamiltonian can be written as:
\begin{equation}
H(\textbf{k}_{\parallel})=\int{\psi_i^{\dagger}(\textbf{k}_{\parallel},k_z)\,\mathcal{H}_{ij}(\textbf{k}_{\parallel},k_z)\,\psi_j(\textbf{k}_{\parallel},k_z)\,\frac{dk_z}{2\pi}}
\end{equation}
where the $6\times 6$ matrix $\mathcal{H}_{ij}(\textbf{k}_{\parallel},k_z)$ is given by Eq.~(\ref{kp}+\ref{Hs}).
Assuming that the $k_z$ dependence is no higher than $k_z^2$ we can write:
\begin{equation}
\mathcal{H}(\textbf{k}_{\parallel},k_z)=\mathcal{H}_0(\textbf{k}_{\parallel})+\mathcal{H}_1(\textbf{k}_{\parallel})\cdot k_z+\mathcal{H}_2(\textbf{k}_{\parallel})\cdot k_z^2
\end{equation} 
where $\mathcal{H}_0,\,\mathcal{H}_1,\,\mathcal{H}_2$ are matrices that do not depend on $k_z$.
By performing a lattice regularization: $k_i\to \frac{1}{a}\sin{(ak_i)}$ and $k_i^2\to \frac{2}{a^2}[1-\cos{(ak_i)}]$, and an inverse Fourier transform to the wave functions only in the $z$-direction: 
\begin{equation}
\psi_j(\textbf{k}_{\parallel},k_z)=\sum\limits_n C_{j,n}(\textbf{k}_{\parallel})\,e^{iak_zn}
\end{equation}
The Hamiltonian becomes:
\begin{align}
H=\sum\limits_{n=1}^N\,&C_{i,n}^{\dagger}(\textbf{k}_{\parallel})\hat{A}_{ij}(\textbf{k}_{\parallel})C_{j,n}(\textbf{k}_{\parallel})+\\ \nonumber
&\Big(C_{i,n}^{\dagger}(\textbf{k}_{\parallel})\hat{T}_{ij}(\textbf{k}_{\parallel})C_{j,n+1}(\textbf{k}_{\parallel})+h.c.\Big)
\end{align}
where $C_{j,n}^{\dagger}(\textbf{k}_{\parallel})$ is the creation operator of a $j$'s band electron in layer $n$ and with in-plane momentum $\textbf{k}_{\parallel}$. $N$ is the number of layers in the $z$-direction and the coefficient matrices are given by $\hat{A}=\mathcal{H}_0+\frac{2}{a^2}\mathcal{H}_2$ and $\hat{T}=\frac{-i}{2a}\mathcal{H}_1-\frac{1}{a^2}\mathcal{H}_2$.
This Hamiltonian can be solved numerically for each value of $\textbf{k}_{\parallel}$ to produce a set of $6N$ eigenvalues and  eigenstates.

\section{Quantum capacitance} \label{qCap}
The quantum capacitance (per unit area) of a two dimensional system is given by, $C_Q=e^2g(\mu)$, where $e$ is the electron charge and $g(\mu)$ is the density of states at the Fermi level. Assuming a Dirac spectrum (with Fermi-velocity $v_F$) for the surface states, the quantum capacitance of a single surface state is given by:
\begin{equation}
C_Q^{sur}=\frac{e^2k_\texttt{F}}{2\pi\hbar v_F}
\end{equation}
where $k_\texttt{F}$ is the Fermi-wave vector which is related to the surface density by $k_F^2=4\pi n$. The geometrical capacitance of the sample is dominated by the oxide layer. Its capacitance is given by: 
\begin{equation}
C_{ox}=\frac{\epsilon_0\epsilon_{ox}}{d_{ox}}
\end{equation}
where $\epsilon_{ox}$ and $d_{ox}$ are the relative permittivity and the thickness of the oxide layer, and $\epsilon_0$ is the vacuum permittivity.
The ratio between these two capacitances is:
\begin{equation}
\frac{C_{ox}}{C_Q^{sur}}=\frac{2\pi\epsilon_0\hbar c}{e^2}\cdot\frac{v_F}{c}\cdot\frac{\epsilon_{ox}}{k_\texttt{F} d_{ox}}=\frac{1}{2\alpha}\frac{v_F}{c}\frac{\epsilon_{ox}}{k_\texttt{F} d_{ox}}
\end{equation}
where $c$ is the speed of light and $\alpha\sim(137)^{-1}$ is the fine structure constant. For HgTe $v_F/c\sim 10^{-3}$, and for typical silicon oxides $\epsilon_{ox}\sim 3$. Using this:
\begin{equation}\label{cond}
\frac{C_{ox}}{C_Q^{sur}}\sim\frac{1}{5k_\texttt{F} d_{ox}}
\end{equation}
As long as $C_{ox}<<C_Q^{sur}$, the full capacitance is given by: $C^{-1}=C_{Q}^{-1}+C_{ox}^{-1}\approx C_{ox}^{-1}$ and the approximation is justified.
Using Eq.~(\ref{cond}), the condition $C_{ox}<<C_Q^{sur}$ is equivalent to:
\begin{equation}
(10\sqrt{\pi}d_{ox})^{-2}<<n
\end{equation}
For typical samples $d_{ox}\sim 110\cdot10^{-7}\,[cm]$, which corresponds to the condition $n>>2.6\cdot10^{7}\,[cm^{-2}]$. In all the calculations we performed, the surface density is in the  range $6-10\,[10^{10}cm^{-2}]$, hence $C_{ox}<<C_Q^{sur}$ and the approximation is justified.

\end{document}